\documentclass{ws-ijmpa}
\usepackage[super,compress]{cite}
\usepackage{graphicx}
\usepackage{slashed}
\usepackage{amsmath}

\newcommand\bef{\begin{figure}}
\newcommand\eef[1]{\label{fg:#1}\end{figure}}
\newcommand\beq{\begin{equation}}
\newcommand\eeq[1]{\label{#1}\end{equation}}
\newcommand\beqa{\begin{eqnarray}}
\newcommand\eeqa[1]{\label{#1}\end{eqnarray}}

\newcommand\fgn[1]{Figure \ref{fg:#1}}
\newcommand\eqn[1]{eq.\ (\ref{#1})}

\newcommand\ie{{\sl i.e.\/}}

\newcommand{\D}{{\cal D}}
\newcommand\m{\mathfrak{m}}
\newcommand{\N}{{\cal N}}
\newcommand\ppbar{\langle\overline\psi\psi\rangle}
\newcommand\tco{T_{co}}

\newcommand{\bilin}[1]{\overline\psi{#1}\psi}

\begin{document}

\title{Real time warm pions from the lattice using an effective theory}
\author{Sourendu\ Gupta}
\address{Department of Theoretical Physics, Tata Institute of Fundamental
         Research,\\ Homi Bhabha Road, Mumbai 400005, India.\\
         sgupta@theory.tifr.res.in}
\author{Rishi\ Sharma}
\address{Department of Theoretical Physics, Tata Institute of Fundamental
         Research,\\ Homi Bhabha Road, Mumbai 400005, India.\\
         rishi@theory.tifr.res.in}
\maketitle

\begin{abstract}
Lattice measurements provide adequate information to fix the parameters
of long distance effective field theories in Euclidean time. Using such a
theory, we examine the analytic continuation of long distance correlation
functions of composite operators at finite temperature from Euclidean to
Minkowski space time. We show through an explicit computation that the
analytic continuation of the pion correlation function is possible and
gives rise to non-trivial effects. Among them is the possibility, supported
by lattice computations of Euclidean correlators, that long distance
excitations can be understood in terms of (very massive) pions even at
temperatures higher than the QCD cross over temperature.\\

\end{abstract}

\section{Introduction}

The computation of the thermodynamics of quantum field theories is under
good control using the Euclidean formulation and non-perturbative lattice
computations. Nevertheless the analytic continuation to real time Minkowski
quantities remains an open problem, despite decades of attempts to chip
away at it. The first attempt to extract a transport coefficient from
lattice computations was made more than three decades ago \cite{wyld}.
However, it wasn't until fifteen years later that it was realized
that more control was needed on the non-perturbative structure of the
spectral density function \cite{gert}. Despite advances in weak-coupling
expansions \cite{amy}, the introduction of new methods \cite{st-as,
st-gu, st-ki}, and many lattice computations \cite{m-sa, m-ka, m-as,
m-da, m-me, m-ph}, the extraction of real-time dynamics at finite
temperature from lattice computations is far from becoming a routine
measurement. Although some spectacular results have been obtained, this
lack of a deeper understanding is a matter of concern.  This is partly
because there have been improved measurements of many flow variables
in heavy-ion collision experiments \cite{flow1, flow2}, and it seems
possible to start on the extraction of transport coefficients from data.

While much attention has been focused on the analytic continuation of
composite operators which lead to transport coefficients, relatively
little has been done about the closely related problem of the analytic
continuation of hadron correlation functions (with the exception
of heavy quarkonia). Hadrons are composite operators in QCD, and at
finite temperature, their Euclidean correlators cannot be connected to
Minkowski correlation functions through reflection positivity. Instead
of focusing on the most general problem of analytic continuation,
our specific discussion will focus on pion properties in real time at
finite temperature.

When discussing thermal effects in QCD, it is useful to keep in mind a
nomenclature introduced by Smilga \cite{Smilga:1996cm}. At a temperature
above the region of the QCD crossover, \ie, for $T\gg\tco$, matter can
be considered hot.  Well below the crossover, \ie, for $T\ll\tco$, he
introduced the term lukewarm matter. The region not covered by either
name, $T\simeq\tco$ we will call warm in this work. We expect that
long-distance properties of lukewarm matter should be described in
terms of hadrons. Chiral symmetry is broken only by quark masses in
hot matter, where hadrons are not expected to be present. Since QCD has
a crossover, rather than a sharp phase transition, it is possible that
warm matter could be described in terms of either hadrons or quarks,
and the more appropriate description is the one which is simpler for
any given quantity.

It may be possible, for example, that the correlation function of flavour
octet axial currents, at distances much longer than $1/T$, is more easily
described in terms of interacting hadron degrees of freedom even for $T$ which
is a little larger than $\tco$.  Of course this description should become
untenable at some $T>\tco$. At the same time, it is possible that the
equation of state of matter, which is dominated by particles of momentum
around $T$, is more easily described by an interacting quark-gluon liquid.
Again, it is possible that this computation remains feasible for some $T$
slightly less than $\tco$ but fails badly for a slightly lower temperature.
If this happens, then there is nothing contradictory about
either description. They could just reflect the complexity of QCD. Rather,
tests of such pictures could be in what happens when the quark masses
vanish, because then the cross over becomes a phase transition.

At zero temperature there are many ways to capture the physics of chiral
symmetry breaking in QCD. At the longest length scales, current algebras
relate pion properties to the (broken) Ward identity of chiral charge
conservation. Chiral perturbation theory is a systematic effective
field theory (EFT) approach that captures an immense amount of hadronic
physics at shorter length scales, while also being compatible with
current algebra. The non-linear sigma model is another such EFT which
captures significant parts of the physics.  However, models such as the
Nambu-Jona-Lasinio (NJL) model also realize the current algebras, and
therefore are interesting approximate low-energy picture of QCD. Quantum
hadro-dynamic models try to mimic some aspects of low-energy hadronic
physics, and are useful tools in some contexts.

A very early calculation in lukewarm chiral perturbation theory
\cite{Gasser:1986vb} indicated that the pion mass, $m_\pi$, would rise
with temperature and the pion decay constant, $f_\pi$, would fall. When
pursued beyond the leading order \cite{Toublan:1997rr}, it became clear
that the definition of a mass was ambiguous. This led to the notion of a
pole mass of a Minkowski propagator, \ie, the energy at which the inverse
pion propagator vanishes.  An ambiguity also arises in the definition of
$f_\pi$. When resolved appropriately, the results of the leading order
computation could be extended, and qualitative changes in the behaviour of
$m_\pi$ and $f_\pi$ were seen.

The earliest computation of lukewarm and warm pion properties in the NJL
model with two flavours of light quarks \cite{Hatsuda:1986gu} obtained
Euclidean pion correlators from a one-loop random phase approximation
analysis of the Bethe-Salpeter equation.  They found that the pion pole
mass increases with temperature, and concluded that this was consistent
with the behaviour of the screening mass measured on the lattice. Later,
a similar computation in the real time formalism \cite{Lutz:1992dv}
for the three flavour NJL model found that the pole mass moved up
with temperature. Both these computations were done in the cutoff
regularization, with parameters fixed to pion properties at $T=0$. The
second paper made a crucial observation: that predictions can be given
in terms of physical quantities by eliminating the cutoff between various
predictions in a given model. An NLO computation in the 1/$N_c$ expansion
of the two-flavour NJL model \cite{Muller:2010am} (taken at $N_c=3$) is
the only such computation to report the pion spectral function. Several
good reviews exist about the NJL model and its application to physics
at finite temperature and density \cite{Vogl:1991qt, Klevansky:1992qe,
Hatsuda:1994pi}

Another set of approaches has been purely phenomenological, utilizing
Lagrangians which express the couplings between many different hadrons.
We will subsume all these approaches under the name of Quantum Hadron
Dynamics (QHD). A very early computation of pion properties in lukewarm
matter \cite{Song:1993ipa} was done in a model with pions coupled to
vector and axial vector mesons with constraints from chiral symmetry for
two flavours. The axial anomaly was taken care of, and a full real time
computation was done to one-loop order. This work made a very clear and
modern analysis of the problem, and reported that screening masses, pole
masses, and the dispersion relation barely change in the lukewarm region.

To the best of our knowledge, there is no previous work which specifically
addressed the differences between pole and screening masses in the
warm region of QCD. This is now accomplished through dual descriptions:
both in terms of quarks and pions. Here we report on an EFT model which
is specifically engineered for finite temperature and has been seen to
successfully reproduce lattice data \cite{Gupta:2017gbs}. We give first
results on how such an effective field theory can be used to analytically
continue Euclidean correlators to real time in the vicinity of the chiral
cross over of QCD.

\section{The effective field theory model}

An EFT is defined with a UV cutoff, $\Lambda$, in the momentum, and
global symmetries play a major role. All local operators, allowed by
symmetry, with mass dimension up to some order $D$ are included in the EFT
Lagrangian. This EFT then should be able to describe the physics of the
system for momenta $k\ll\Lambda$ with accuracy $(k/\Lambda)^D$. Because
the number of terms is potentially large, so is the number of couplings
in the EFT. They are fixed by requiring the EFT to reproduce a sufficient
number of measurements. After choosing which measurement to match, if
one varies $\Lambda$ by a small amount, then the values of the fitted
parameters also change by a small amount. This is to be interpreted
as the analogue of RG running in this cutoff field theory. Once the
parameters are fixed, everything else is a prediction of the EFT.

The existence of a special frame where the heat bath is at rest
implies a lack of boost invariance in the Lagrangian, as a result
of which the global symmetries are of spatial rotations, apart from
a SU(2)$\times$SU(2) chiral-flavour symmetry.  We consider a model
\cite{Gupta:2017gbs} of self-interacting quarks, the Minkowski version
of which has the Lagrangian
\beqa
\nonumber
  L &=& -d^3T_0\bilin{} + \bilin{\slashed\partial^0} 
      + d^4 \bilin{\slashed\nabla} 
      - \frac{d^{61}}{T_0^2} \left[(\bilin{})^2 
            + (\bilin{i\gamma_5\tau^a})^2 \right]
\\ \nonumber &&
      - \frac{d^{62}}{T_0^2} \left[(\bilin{\tau^a})^2 
            + (\bilin{i\gamma_5})^2 \right] 
      - \frac{d^{63}}{T_0^2} (\bilin{\gamma_0})(\bilin{\gamma^0}) 
      - \frac{d^{64}}{T_0^2} (\bilin{i\gamma_i})(\bilin{i\gamma^i})
\\ \nonumber &&
      - \frac{d^{65}}{T_0^2} (\bilin{\gamma_5\gamma_0})
                             (\bilin{\gamma_5\gamma^0}) 
      - \frac{d^{66}}{T_0^2} (\bilin{i\gamma_5\gamma_i})
                             (\bilin{i\gamma_5\gamma^i})  
\\ \nonumber &&
      - \frac{d^{67}}{T_0^2} \left[
      (\bilin{\gamma_0\tau^a})
      (\bilin{\gamma^0\tau^a}) 
      + (\bilin{\gamma_5\gamma_0\tau^a})
        (\bilin{\gamma_5\gamma^0\tau^a})\right]
\\ \nonumber &&
      - \frac{d^{68}}{T_0^2} \left[
              (\bilin{i\gamma_i\tau^a})(\bilin{i\gamma^i\tau^a}) 
            + (\bilin{i\gamma_5\gamma_i\tau^a})(\bilin{i\gamma_5\gamma^i\tau^a})\right]  
\\ \nonumber &&
- \frac{d^{69}}{T_0^2} \left[(\bilin{iS_{i0}})(\bilin{iS^{i0}}) 
            + (\bilin{S_{ij}\tau^a})^2 \right]
\\ &&
      - \frac{d^{60}}{T_0^2} \left[(\bilin{iS_{i0}\tau^a})(\bilin{iS^{i0}\tau^a}) 
            + (\bilin{S_{ij}})^2 \right].
\eeqa{qlag}
where $\slashed\partial^0=\gamma_0\partial^0$, also $\slashed\nabla =
\gamma_j\partial^j$ with $j$ running over all spatial indices, and $\tau^a$
are the generators of flavour SU(2). We use Weinberg's metric conventions
\cite{wein}. This theory is defined with a cutoff, $T_0$, which we will
choose so that physics at the temperatures of interest can be described
by the theory.  For later simplicity in writing formul{\ae}, we introduce
the notation $\N=4N_cN_f$ for the number of components of quark fields
and $m_0=d^3T_0$.

Since the EFT is similar to the Nambu-Jona-Lasinio (NJL) model \cite{njl},
known techniques \cite{Klevansky:1992qe} can be used to first analyze
the mean field theory. In the chiral limit, $m_0=0$, chiral symmetry
is broken spontaneously, a non-vanishing quark condensate, $\ppbar$, is
produced, and a second-order chiral symmetry restoring phase transition
(at a temperature that we choose to be $T_0$) is found. The single
combination of the dimension-6 couplings,
\beq
 \lambda=(\N+2)d^{61}-2d^{62}-d^{63}+3d^{64}+d^{65}-3d^{66}
    -\frac32d^{69}+\frac32d^{60},
\eeq{lambda}
appears in the subsequent physics that we discuss \footnote{This
formula corrects an earlier typographical error \cite{Gupta:2017gbs}.}. In the
Euclidean theory these conclusions followed from the computation of the
free energy. In real time they come from a self-consistent solution of
the one-loop Dyson-Schwinger equation for the quark propagator
using real time perturbation theory 
\cite{Kobes:1984vb, Kobes:1985}. In both cases we use dimensional
regularization with $\overline{MS}$ subtraction. Since the expressions
for the quark condensate are exactly the same in the two computations,
the phase structure of the theory can be computed in either the Euclidean
or real-time formalism \cite{Landsman:1986uw}.

\bef
\begin{center}
\includegraphics[scale=0.45]{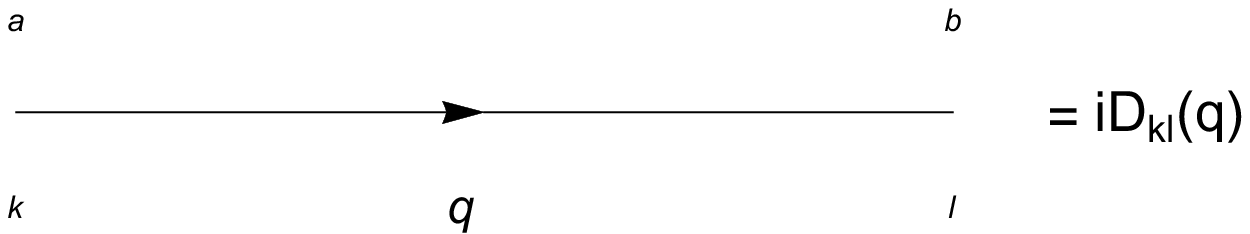}\hfill
\includegraphics[scale=0.45]{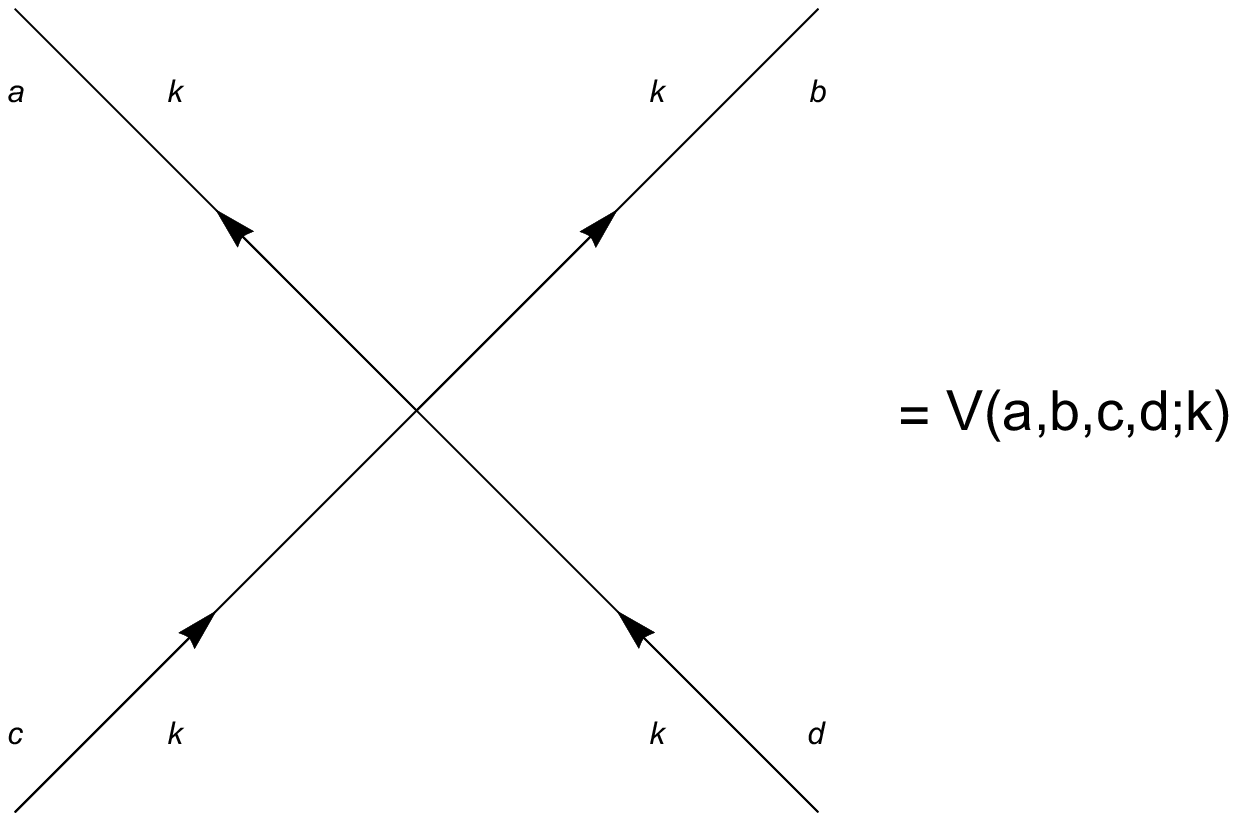}

\bigskip

\includegraphics[scale=1.00]{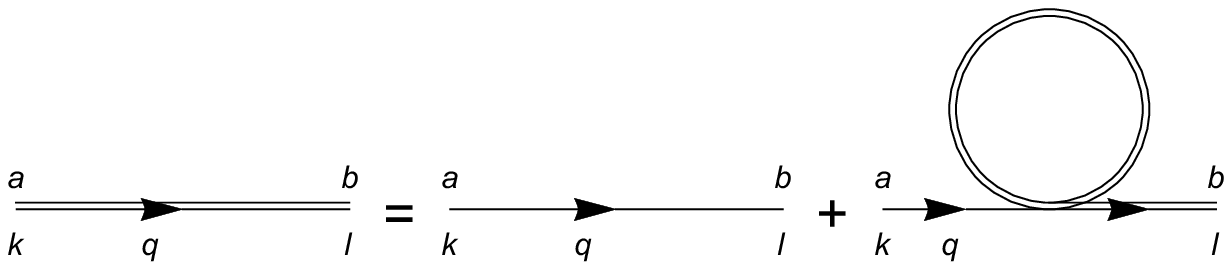}
\end{center}
\caption{The Feynman rules for the quark EFT in \eqn{qlag}. The thermal
 propagator $D(q)$ is the $2\times2$ matrix given in \cite{Kobes:1985}.
 The indices $k$ and $l$ are thermal indices which take values 1 and 2,
 the rest are composite Dirac, flavour, colour indices.
 The vertex is diagonal in the thermal index $k$ (which takes values 1
 and 2), with $V(a,b,c,d;k)=(-1)^k(2id^{6A}/T_0^2)\Gamma^A_{ac}\Gamma^A_{bd}$,
 where $A$ runs over all the operator types listed in \eqn{qlag}.
 A diagrammatic representation is also given of the one-loop Dyson-Schwinger
 equation for the real-time Fermion propagator arising from the
 Lagrangian in \eqn{qlag}.}
\eef{quarkrules}

The Euclidean computation has been given in detail in \cite{Gupta:2017gbs}.
Here we outline how the real time computation proceeds. The Feynman rules
for the EFT are shown in \fgn{quarkrules}. Note the fact that the vertex
is diagonal in all the thermal indices. Denoting the resummed Dyson-Schwinger propagator as $i{\cal D}(q)$, the Neumann series corresponding to
the diagrammatic equation in \fgn{quarkrules} can be written as
\beq
 i{\cal D}(q)=iD(q)+iD(q)\Sigma{\cal D}(q)
\eeq{dyschw}
The very simple structure of the vertex factor then allows us to
decouple the equations for ${\cal D}_{11}$. This forces all the internal
quark lines in the superdaisy graphs to be $D_{11}$. As a result, the
computation reduces to well known pieces \cite{njl, Landsman:1986uw}. The
one-loop self-energy $\sigma$, whose iteration would give the $\Sigma$
in \eqn{dyschw}, is given by
\beq
 \sigma_{lk}=-\sum_A\frac{2id^{6A}}{T_0^2}\int\frac{d^4p}{(2\pi)^4}
  \left[\Gamma^A_{mn}\{D_{11}(p)\}_{nm}\Gamma^A_{lk} 
       -\Gamma^A_{ln}\{D_{11}(p)\}_{nm}\Gamma^A_{mk} \right].
\eeq{selfenergy}
The integral gives zero when the numerator has a linear term in the
momentum $p$. Therefore, the result is diagonal in spinor-flavour-colour
indices. Performing the traces, we get exactly the linear combination
of the couplings shown in \eqn{lambda}. Furthermore, the result is
real. Then putting this into the Dyson-Schwinger resummation, one obtains
a self-consistency equation for the Fermion mass which is exactly the
same as the gap equation in Euclidean space.

At finite $m_0$, the quark mass explicitly breaks chiral
symmetry. Nevertheless, a remnant of spontaneous chiral symmetry breaking
appears as a large value of the condensate, giving an effective quark mass
$m=m_0+\Sigma$, where $\Sigma=2\lambda\ppbar/T_0^2$. As the temperature
is increased, $\Sigma$ crosses over to a small value. This is exactly
what was found in the Euclidean computation \cite{Gupta:2017gbs}.

The fixing of the parameters in the action of \eqn{qlag} was done by
matching one-loop expressions for the axial current correlators to
lattice measurements \cite{brandt}. The process was simple; since the
axial symmetry is broken, small fluctuations around the mean field have
the quantum numbers of the pion. These fluctuations can be parametrized
by the following field redefinition when using a Hubbard-Stratanovich
transformation,
\beq
 \psi\to\exp\left[\frac{i{\bf\pi}\gamma^5}{2f}\right]\,\psi
\eeq{hst}
with a three component field ${\bf\pi}=\pi^a\tau^a$,
and a constant $f$ with the dimension of mass.  We will show elsewhere
that this transformation correctly takes into account exchange effects
which capture the correct value of $\lambda$ given in \eqn{lambda}.
Introducing this parametrization into the quark action and expanding
to second order in $\pi$ gives a coupled model of quarks and mesons
\beq
 L_c = -\bilin{\left[d^3T_0\left(1+i\gamma^5\frac\pi f
           -\frac{\pi^2}{2f^2}\right)-\slashed\partial
           -\frac i{2f}\gamma^5\slashed\partial\pi\right]} + \cdots
\eeq{coupled}
where $\slashed\partial=\slashed\partial^0+d^4\slashed\nabla$, and the
terms of dimension-6 and higher have not been written out.  The pion
appears as an auxiliary field, and hence has no kinetic term in $L_c$.
After integrating over the quark fields, an effective pion action
was obtained, with parameters of the action expressed in terms of the
parameters in the quark Lagrangian.

In the chiral limit the axial vector current is conserved; the
conservation law is broken only by the parameter $d^3$ in the action.
Using the one-loop version of the PCAC relation at finite temperature,
the long-distance static axial current correlator could be parametrized
in terms of the coupling constants $d^3$ and $d^4$. The convention
that $T_0$ is the chiral transition temperature $T_c$ fixes $\lambda$,
and the value of $T_0$ was then inferred from matching the cross over
temperature, $\tco$, at finite $m_0$ simultaneously with the matching
of axial current correlators \cite{Gupta:2017gbs}.

\section{Analytic continuation of correlation functions}

The specific question that we ask here is how the parameters which
govern the long-distance part of the pion correlation function can be
analytically continued to real time. Since we have a EFT which is written
in terms of the pions, it should be a simple matter to use it to do the
continuation. We present this here.

\bef
\begin{center}
\includegraphics[scale=0.45]{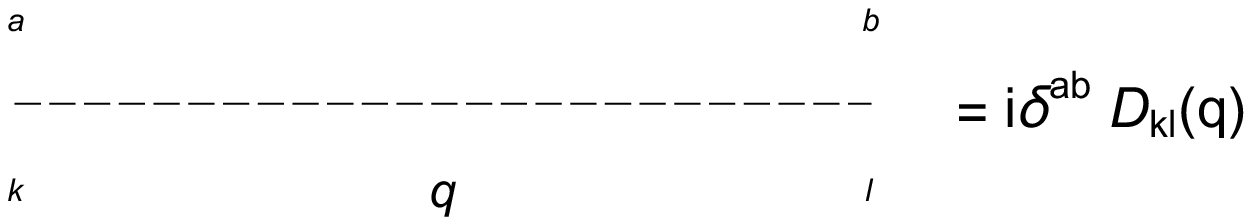} \hfill
\includegraphics[scale=0.45]{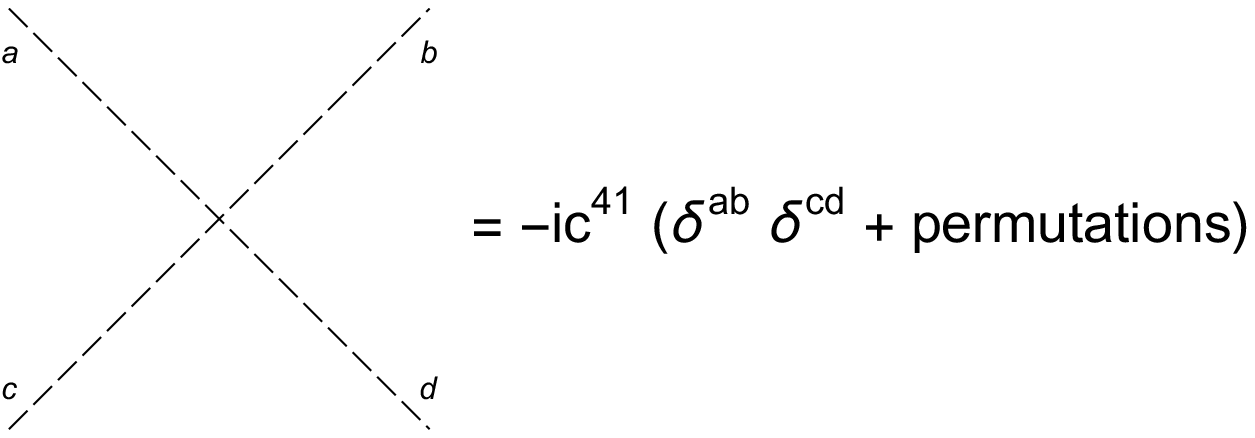}
\end{center}
\caption{The Feynman rules for pions generated by the Lagrangian in
 \eqn{pilag}. The flavour indices are $a,b,c,d$. The indices $k$ and $l$
 which run over 1 and 2 refer to the thermal components of fields.}
\eef{pifeyn}

The Minkowski Lagrangian for the pion EFT has the form
\beq
  L_f = -\frac12c^2 T_0^2\pi^2 + \frac12(\partial_0\pi)^2 
    - \frac12c^4(\nabla\pi)^2 - \frac{c^{41}}8\pi^4.
\eeq{pilag}
which includes all terms of dimension up to 4. In the rest of this
paper we will assume that all the dimensionless coupling parameters have
exactly the value that they have with the one-loop matching described
in the previous section, and use the shorthand notation $\m^2=c^2
T_0^2$. The pole mass computed to higher loop order in the pion theory
will be referred to as $m_\pi$.

We need to compute only the 11 component of the $2\times2$
thermal correlation function 
because we are interested in external
pion states \cite{Kobes:1984vb}.
At tree level the pion correlation function is give by
\beq
  \D^\pi_{11}(q) = \left[q_0^2-c^4|{\bf q}|^2-\m^2+i\epsilon\right]^{-1}
   + 2\pi in_B(q_0)\,\delta\left(q_0^2-c^4|{\bf q}|^2-\m^2\right),
\eeq{treelevel}
where $n_B$ denotes the Bose distribution function. The spectral function,
\beq
  \rho(q^0,|{\bf q}|) = i D_R(q^0,|{\bf q}|) - i D_A(q^0,|{\bf q}|),
\eeq{specfn}
which is the imaginary part of the causal correlator, has support only
on the ``acoustic shell'' $q_0^2=c^4|{\bf q}|^2-\m^2$. This simple
spectral function means that there is no special subtlety in analytic
continuation from spacelike to timelike momenta.  Since the spacelike
region is captured in Euclidean lattice computations, this means that
the values of $c^2$, $c^4$, and $c^{41}$ are those extracted by matching
lattice data \cite{Gupta:2017gbs}.

\bef[tbh]
\begin{center}
\includegraphics[scale=0.35]{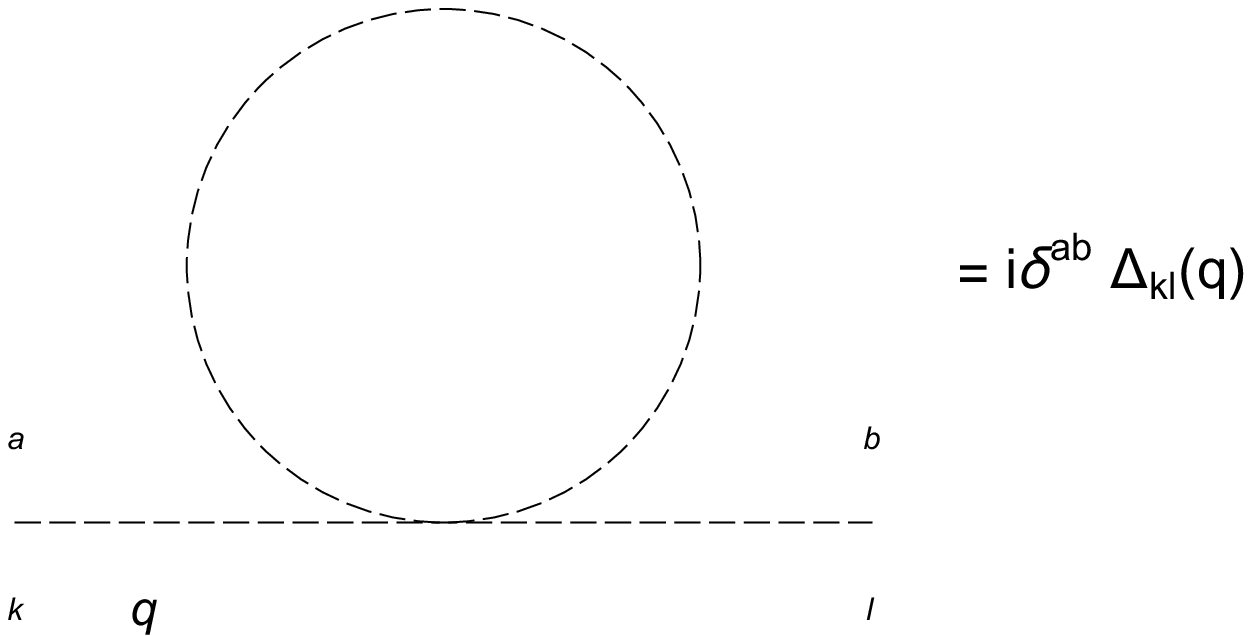}
\end{center}
\caption{The Feynman diagram for the pion correlation function up to
 one loop order. The external momentum is $q$, and the external flavour
 indices are $a,b$.}
\eef{pi1loop}

The Feynman rules for the pion theory are given in \fgn{pifeyn}. Note
the simple form of the vertex which was also pointed out earlier
for the quark computation; the vertex forces all fields connected to
it to be either type 1 or type 2. The one-loop correction is given in
\fgn{pi1loop}. The cutting rules \cite{Kobes:1984vb} at finite temperature
can be understood as dividing vertices into two categories, circled and
uncircled, and cutting the propagators which connect two different types
of vertices. Since there is only one vertex in this diagram, no cuts
are possible.  As a result, the diagram gives a purely real contribution.
The flavour indices at the vertex contract in two ways. The contraction
$\delta^{ab}\delta^{cc}$ gives $(N_f^2-1)/2$, and $\delta^{ac}\delta^{cb}$
gives unity.  As a result, the one-loop self-energy can be written as
\beq
 \delta m_\pi^2 = \frac i2c^{41}(1+N_f^2)
      \int\frac{d^4k}{(2\pi)^4} D^\pi_{11}(k).
\eeq{oneloop} 
Since this is independent of $q$, it gives no contribution to $c^4$,
$c^{41}$ or $f$.  With the $\overline{{\rm MS}}$ regularization
\eqn{oneloop} gives
\beq
 \delta m_\pi^2 = \frac{c^{41}}{2(c^4)^{3/2}}(1+N_f^2) \left[
  \frac{\m^2}{16\pi^2}\left\{\log\left(\frac{\m^2}{c^4M^2}\right)
   -1\right\}
      +\int\frac{d^3k}{E_k(2\pi)^3} n_B(E_k)\right],
\eeq{msbaroneloop} 
where we have scaled $c^4$ out of the integral, and then defined
$E_k=\sqrt{k^2+\m^2}$.  The choice of the renormalization point,
$M$, is arbitrary; we will choose it to be the same as in the quark
theory. The pole mass is then $m_\pi^2=\m^2+\delta m_\pi^2$. Proceeding
to the Dyson-Schwinger superdaisy resummation \cite{Dolan:1973qd,
Klevansky:1992qe} is straightforward and follows the same lines as in
the quark case. There are no complications due to spinor indices, and
the resummation of the result in \eqn{msbaroneloop} is straightforward;
one replaces $\m^2$ on the right hand side by $\m^2+\delta_\pi^2$ and
solves the equation numerically for $\delta m_\pi$.

\bef[thb]
\begin{center}
 \includegraphics[scale=0.80]{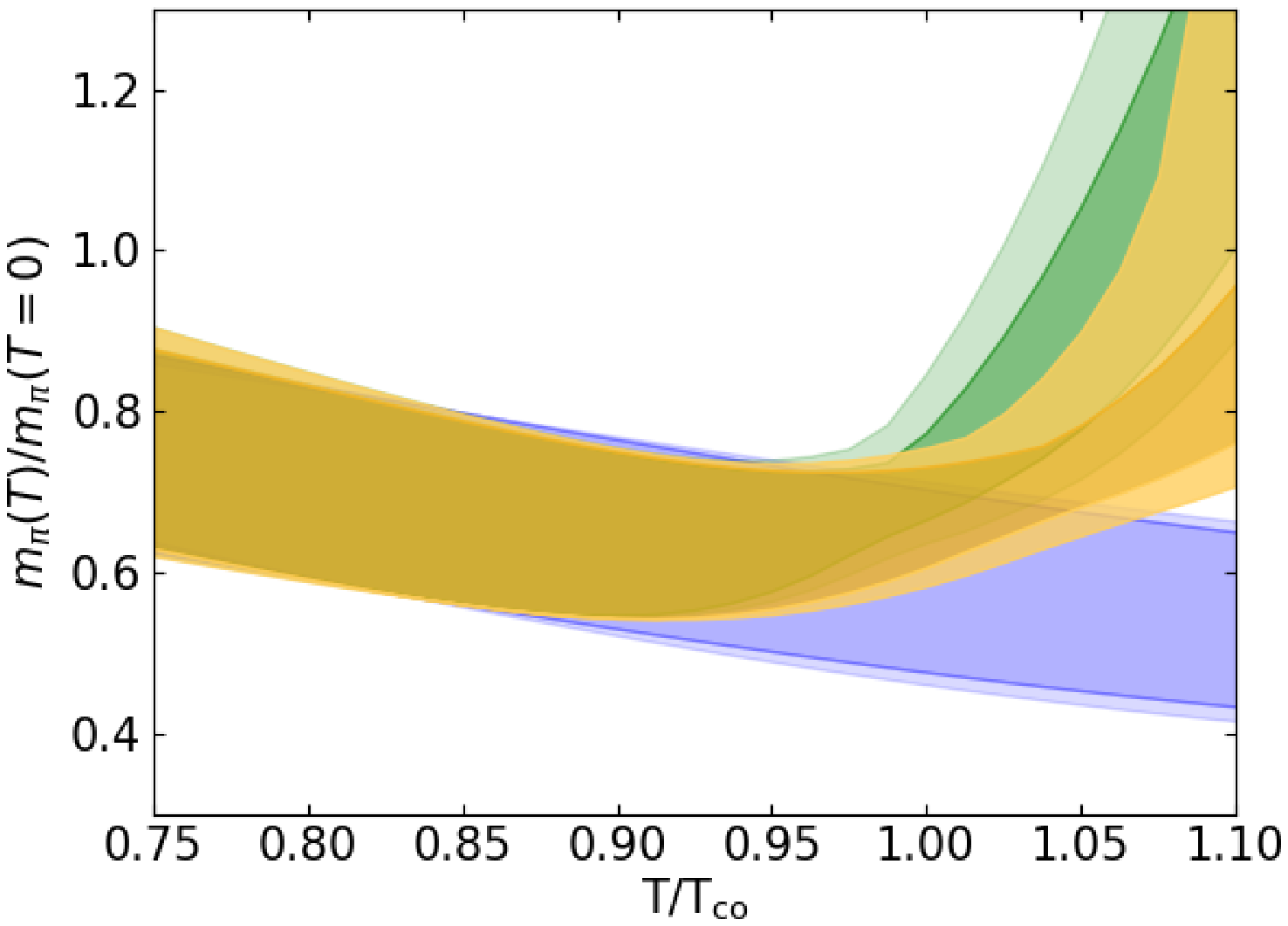}
\end{center}
\caption{The pion pole mass as a function of temperature. In warm
 QCD, the pion is strongly affected by temperature. The band in blue
 shows the tree level, in green the one loop corrected, and in gold the
 one loop Dyson-Schwinger resummed result. The darker inner part of
 each band is the region allowed by the statistical uncertainties in
 parameters \cite{Gupta:2017gbs} for $M=1.6\pi T_0$, and the lighter
 outer part is the result of a variation in the renormalization scale
 $1.2\le M/(\pi T_0)\le2.0$. The pion mass at $T=0$ is taken from a
 lattice computation with the same Lagrangian and parameters as the
 finite temperature computation to which the matching is performed.}
\eef{polemass}

The results for the pole mass are shown in \fgn{polemass}. One sees
that in the whole warm region of QCD, thermal effects are important,
since the pion pole mass differs significantly from the pion mass at zero
temperature, and is actually smaller. We note that the pole mass rises as
the temperature approaches $T_{co}$.  We have continued the computation
to temperatures slightly above $T_{co}$, because, as discussed earlier,
a cross over does not necessarily prevent a hadronic description from
holding above this temperature. Whether this is in agreement with QCD
is discussed next.

\bef[thb]
\begin{center}
 \includegraphics[scale=0.38]{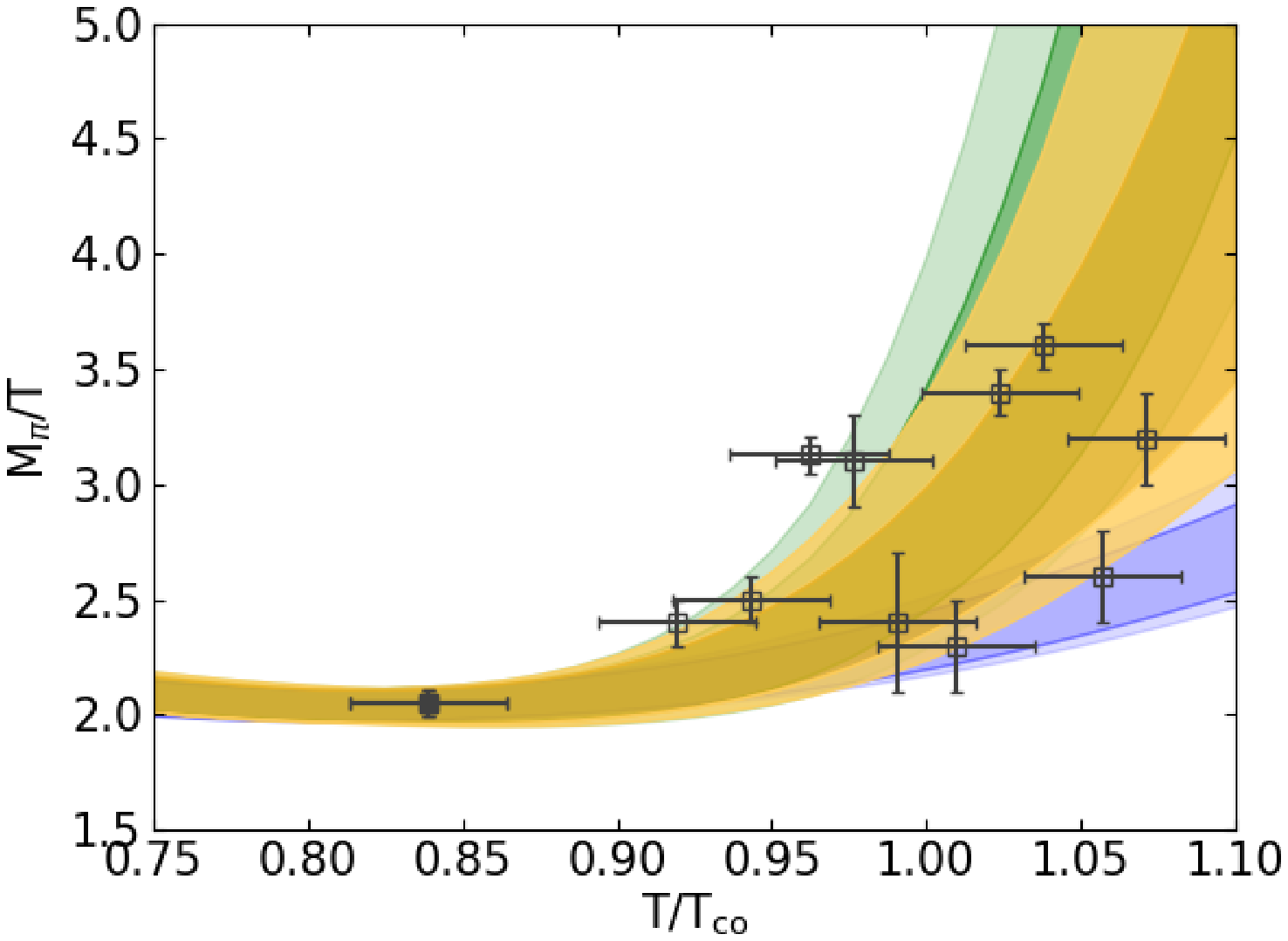}
 \includegraphics[scale=0.38]{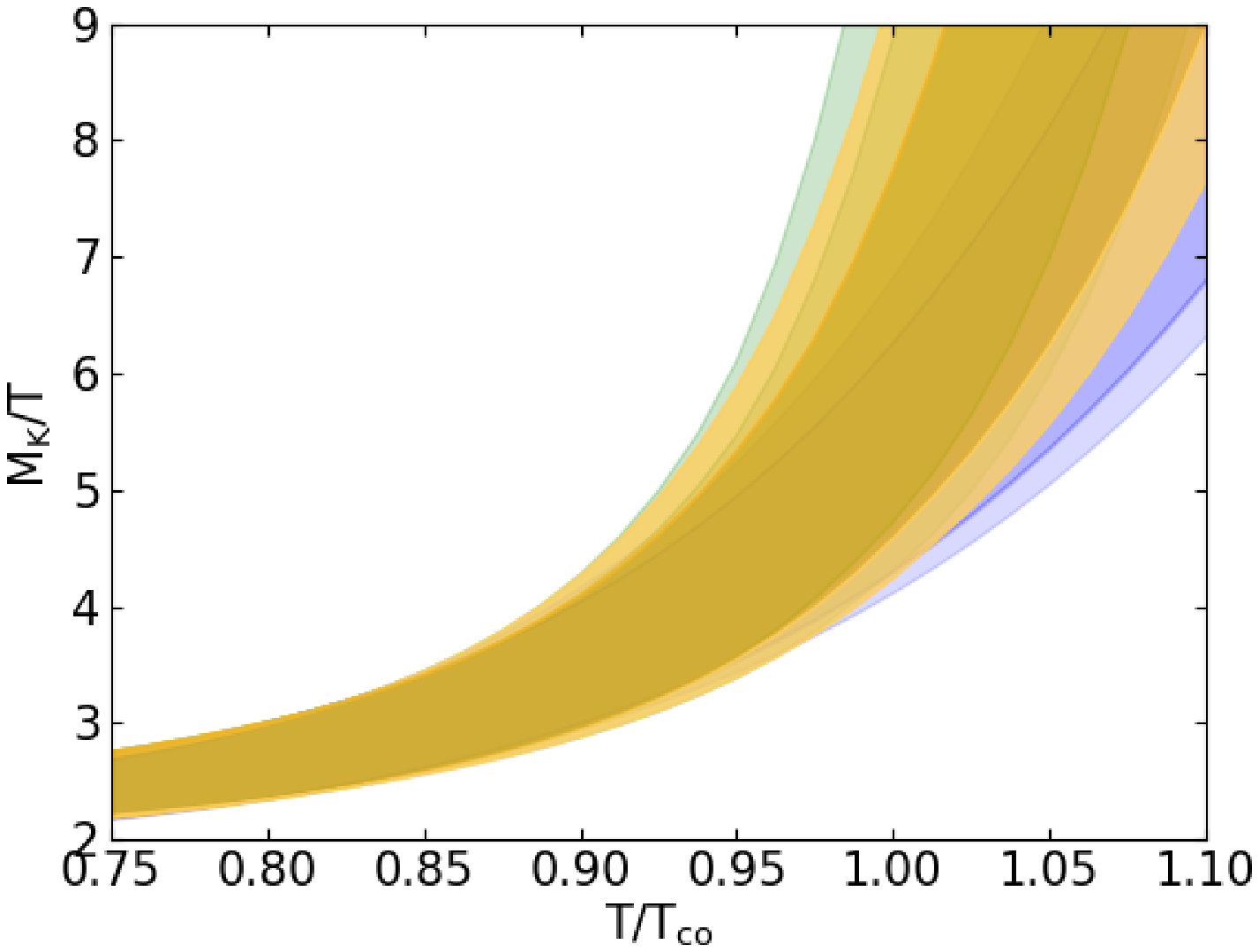}
\end{center}
\caption{The pion screening mass (left) and kinetic mass (right). The
 band in blue shows the tree level, in green the one loop corrected,
 and in gold the one loop Dyson-Schwinger resummed result. The darker
 inner part of each band is the result of the statistical uncertainties
 in parameters \cite{Gupta:2017gbs}, and the lighter outer part is
 the result of a variation in $M/T_0$. The lattice measurements of the
 screening mass \cite{brandt} are also shown. The filled point is the
 only screening mass which was used in the extraction of EFT parameters;
 the rest are predictions.}
\eef{results}

Static correlators fall off with distance exponentially with the
Debye screening mass, $M_\pi=m_\pi/\sqrt{c^4}$. For real-time dynamic
correlators, the dispersion relation for small $q$ is $E=m_\pi+|{\bf
q}|^2/(2 M_K)$, where the rest mass is exactly the pole mass,
but the kinetic energy involves a kinetic mass, which is given
by $M_K=m_\pi/c^4$. Results for both are shown in \fgn{results}.
Note that the one-loop result gives an improved description of the
lattice measurements of $M_\pi$ \cite{brandt} even at temperatures a
little above $\tco$. We note that to this order, the appropriate values of
$c^4$ and $f$ remain as obtained from Euclidean lattice computations. We
have more remarks on this in the next section.

\section{Issues}

The agreement of screening masses measured on the lattice with the
one loop values raises several interesting physics questions. Since
the pole mass of the pion at finite temperature is smaller than the
pion mass at $T=0$ (see \fgn{polemass}) earlier computations in chiral
perturbation theory \cite{Gasser:1986vb,Toublan:1997rr} and the NJL
model \cite{Hatsuda:1986gu,Lutz:1992dv} must differ in some piece of
the physics. We believe it is the fact that the breaking of Lorentz
invariance allows the pion kinetic term to take on a non-trivial value
of $c^4$. This is constrained by lattice data, and gives a difference
between the pole and screening masses.  Since it varies extremely rapidly
in the warm region, it is important. From \fgn{polemass} it seems that
$m_\pi(T)/m_\pi(T=0)$ could become closer to unity at smaller temperature.
Whether or not it exceeds unity at some intermediate temperature in the
lukewarm region remains an open question.

The other question is even more intriguing. Until what temperature
can the pion theory be used? Pion fields were introduced to capture
small fluctuations around a local minimum of the free energy. In
the chiral limit this can no longer make sense beyond $T_c$. When
there is a cross over, the symmetry broken minimum does not vanish
abruptly. However, at some temperature the ``radial'' fluctuations,
corresponding to a scalar mode, can no longer be neglected, and the pure
pion theory cannot make sense. It has been seen in lattice computations
\cite{Gupta:2013vha,Bazavov:2019www} that pion and scalar masses remain
distinct at temperatures lower than $1.1\tco$. This could be the reason
why the pion theory seems to explain physics at temperature higher than
$\tco$, albeit with a pion pole mass which rapidly increases. There is
a range of temperatures where one has chiral symmetry restoration but
long-distance correlations are still mediated by a hadronic excitation,
namely a very massive pion.

One could turn the question around, and ask whether the compositeness of
the pion should be expected show up in the computation of correlation
functions at long-distances. A computation in which compositeness was
assumed to show up was reported long back in the large $N$ Gross-Neveu
model at finite temperature \cite{precursors}. It was also reported
that similar results are obtained when pion correlators are computed
using the quark EFT to one loop \cite{Gupta:2019qri}, except that the
Gell-Mann-Oakes-Renner relation was satisfied (the difference could
be due to the fact that a $\overline{\rm MS}$ regularization, which
preserves symmetries, was used in the EFT computation). However, there
is ample observational evidence in a variety of physical systems that
long-distance correlation functions of Goldstone bosons do not see the
underlying compositeness. The most straightforward experimental evidence
for this is in Andreev reflections: if an electron is incident on a
normal-superconducting boundary, a hole is reflected from the interface,
and only a Cooper pair propagates inside the superconductor. This
happens not only at $T=0$ but also at finite temperature, as long as the
superconducting gap is large enough. There is no violation of decoupling
theorems. What this means for QCD is that one should expect that the
long-distance correlators of pions can be computed in the pion EFT, as
long as the pion remains the lightest mode around the symmetry-broken
minimum.

Nevertheless, there is some interesting, and generic, physics which is not
captured in the computation presented here.  A collection of particles
in a heat bath scatter off each other, giving rise to quite non-trivial
spectral functions. This is not seen in the one loop result, for the
simple reason that the external momentum does not enter the loop. It is
clear that two loop computations will give more interesting, and generic,
structures. In particular, the dependence on external momentum will
modify all couplings as well as $f$, and the corrections could differ
in the spacelike and timelike regimes. Such a computation is currently
underway, and the results will be reported soon.

\end{document}